\documentclass[preprint,12pt,authoryear]{elsarticle}
\usepackage{graphicx}
\usepackage{xcolor}
\usepackage{cancel}
\usepackage{natbib}
\usepackage[colorlinks=true, citecolor=blue, urlcolor=blue]{hyperref}
\usepackage{doi}
\usepackage{array, makecell}
\usepackage{rotating}
\usepackage{verbatim}
\usepackage{float}

\journal{Journal of Informetrics}

\begin{document}

\begin{frontmatter}

\title{Big Tech influence over AI research revisited: memetic analysis of attribution of ideas to affiliation}

%\titlerunning{Big Tech influence over AI research revisited}

\author[a]{Stanisław Giziński}
\author[a]{Paulina Kaczyńska}
\author[b]{Hubert Ruczyński}
\author[a,c]{Emilia Wiśnios}
\author[d]{Bartosz Pieliński}
\author[a,b]{Przemysław Biecek}
\author[e]{Julian Sienkiewicz \corref{ca}}
\cortext[ca]{Corresponding author, email: \url{julian.sienkiewicz@pw.edu.pl}}
%\authorrunning{S. Giziński et al.}

\affiliation[a]{organization={University of Warsaw, Faculty of Mathematics, Informatics and Mechanics}, city={Warsaw},country={Poland}}
\affiliation[b]{organization={Warsaw University of Technology, Faculty of Mathematics and Information Science}, city={Warsaw},country={Poland}}
\affiliation[c]{organization={NASK National Research Institute}, city={Warsaw},country={Poland}}
\affiliation[d]{organization={University of Warsaw, Faculty of Political Science and International Studies}, city={Warsaw},country={Poland}}
\affiliation[e]{organization={Warsaw University of Technology, Faculty of Physics}, city={Warsaw},country={Poland}}
%
%\maketitle            

\begin{abstract}
There exists a growing discourse around the domination of Big Tech on the landscape of artificial intelligence (AI) research, yet our comprehension of this phenomenon remains cursory. This paper aims to broaden and deepen our understanding of Big Tech's reach and power within AI research. It~highlights the dominance not merely in terms of sheer publication volume but rather in the propagation of new ideas or \textit{memes}. Current studies often oversimplify the concept of influence to the share of affiliations in academic papers, typically sourced from limited databases such as arXiv or specific academic conferences.

The main goal of this paper is to unravel the specific nuances of such influence, determining which AI ideas are predominantly driven by Big Tech entities. By employing network and memetic analysis on AI-oriented paper abstracts and their citation network, we are able to grasp a deeper insight into this phenomenon. By utilizing two databases: \textit{OpenAlex} and \textit{S2ORC}, we are able to perform such analysis on a much bigger scale than previous attempts.

Our findings suggest that while Big Tech-affiliated papers are disproportionately more cited in some areas, the most cited papers are those affiliated with both Big Tech and Academia. Focusing on the most contagious memes, their attribution to specific affiliation groups (Big Tech, Academia, mixed affiliation) seems equally distributed between those three groups. This suggests that the notion of Big Tech domination over AI research is oversimplified in the discourse.

\end{abstract}

\begin{keyword}Knowledge Diffusion \sep Novelty \sep Affiliation Influence \sep Big Tech Impact \sep Complex Networks \sep Natural Language Processing
\end{keyword}
\end{frontmatter}
\section{Introduction}

Artificial Intelligence (AI) research is often seen as dominated by Big~Tech companies. These companies, with their immense computing resources and access to vast amounts of data, have influenced the development of the field. However, while this influence has been beneficial in certain aspects, it has also raised concerns that the strong position of Big~Tech could influence the direction and character of AI, resulting in significant losses for society and science.
% reproducibility
One of the issues raised is the reproducibility of Big~Tech's research~\citep{ebell_towards_2021}. Given their proprietary datasets and custom software, independently replicating their studies is often difficult. 
% conflict interest
Conflict of interest is another issue that has been raised by researchers~\citep{abdalla_grey_2021, young_confronting_2022, hagendorff_ethical_2021}. One notable case that exposed these concerns was that of Dr. Timnit Gebru, formerly part of Google's AI ethics team. Her dismissal highlighted the conflicts of interest arising when industry-aligned research agendas collide with ethical considerations. 
% privatisation of AI knowledge
There is also evidence that the longer researchers remain in the industry, the more their high-impact work becomes privatized, limiting the dissemination of knowledge that could otherwise benefit the academic community~\citep{jurowetzki_privatization_2021}.

While there are many concerns about potential threats to research integrity due to this influence, it is equally important to recognize the positive impact of Big Techs . First, their computational resources and data, previously inaccessible on such a scale, provide unprecedented opportunities to discover and develop novel Machine Learning (ML) algorithms. In addition, Big Tech--Academia collaboration may benefit scientific endeavors for reasons other than access to resources. \citet{evans_industry_2010}~argues that the industry's relative lack of interest, in theory, could result in their academic partners being more likely to produce novel and theoretically unexpected experiments than academics working by themselves. 

%why analyze this?
The matters mentioned above indicate that the influence of Big Tech (and more broadly the private sector, also referred to as \textit{company} in our paper) on AI research needs to be carefully examined.
% current state of knowledge 
Several such studies in this regard exist, including analysis of Big Tech's funding of academic AI researchers \citep{abdalla_grey_2021}, changes in the proportion of Big~Tech--affiliated papers at top conferences \citep{ahmed_-democratization_2020}, the flow of researchers from Academia to the private sector \citep{jurowetzki_privatization_2021}, the focus of the private sector on specific sub-fields of AI \citep{klinger_narrowing_2020}, and the comparison of values encoded in the papers between the private sector and Academia \citep{birhane_values_2021}.

Unfortunately, approaches to this topic mentioned above are limited. The methods used so far to quantify Big Tech's dominance tend to oversimplify the concept of influence by perceiving it as the share of affiliations in papers regarding a specific topic. Methods are limited to tools such as topic modeling and keyword analysis. Current research also lacks diversity in data sources, relying heavily on preprint servers or specific conferences to collect papers. In the previous studies, the association of papers with affiliations was binary -- each paper was associated with either Big Tech (or the private sector in general) or Academia, depending on the first author or the proportion of affiliations in the paper. This limits possible conclusions, as the Company--Academia collaboration in the area of AI is widespread.

All of the above limitations, combined with strong statements \citep[e.g. researchers on this topic do not shy away from comparing Big Tech to Big Tobacco,][]{abdalla_grey_2021}, could result in a selective view of the matter, leading to the tension between Big Tech and Academia, and losses for science and society. Through our study, we aim to facilitate a more nuanced understanding of Big Tech and Academia dynamics, highlighting strengths and weaknesses within this relationship. We critically evaluate existing methodologies, advocate for more holistic measures of influence, and explore the implications of this dominance for the field of AI to foster further discourse about the symbiosis between Academia and industry.

In this study, along with a network analysis aimed at quantifying the influence of Big~Tech and Academia on the AI papers citation network, we performed memetic analysis. A meme is understood here as a piece of an idea that is transmitted through culture. In the context of research papers, this transmission occurs through citation. Using \textit{meme score}, which quantifies the replicating power of a specific meme, we measure the spread of particular ideas in AI research. In addition, we quantify the probability of particular memes being replicated, conditioned on the affiliation of the authors of the paper containing the meme. This approach allows us to understand how the ``spreading power'' of specific ideas depends on affiliation groups (Big~Tech, Academia, Companies, and mixed), which sheds light on which ideas Big~Tech has more influence on spreading. Moreover, we investigated how papers with joint Big Tech--Academia affiliations differ from papers authored purely by authors affiliated with one category. In summary, we address the following research questions:
\begin{enumerate}
    \item Do papers affiliated with both Big Tech and Academia differ in citation distributions from papers affiliated only with Big Tech or Academia? Do Big Tech affiliations differ from Company ones (i.e., industrial but not attributed to Big Tech)?
    \item What ideas in AI research are the most contagious?
    \item Does the contagiousness of a meme differ depending on the affiliation of paper authors?
    \item What ideas are more contagious when discussed by Big Tech?
\end{enumerate}

\section{Related Work}

\subsection{Big tech influence over AI research}
Several studies have attempted to quantify Big Tech's influence in AI research using a variety of methods and data sources.
% funding
\citet{abdalla_grey_2021} examine the funding provided by Big Tech to academic AI researchers. The study identifies a recurring pattern where private companies increase their support for academic institutions when their public image declines, often due to media incidents such as the Cambridge Analytica scandal. The~authors compare this funding pattern to the tactics employed by the tobacco industry. The study reveals that 59\% of papers published in top journals that address AI's ethical and societal implications include at least one author with financial ties to a Big Tech company.

% papers published at top conferences
\citet{ahmed_-democratization_2020} examine the dynamics of participation in major AI conferences following the rise of deep learning in 2012. The~authors analyze 171,394 papers from 57 computer science conferences. They~find an increase in the participation of large technology companies, particularly since 2012. In addition, the paper uses term frequency analysis of abstracts to uncover distinct research areas among different organizations.

% brain-drain and flow of researchrs
\citet{jurowetzki_privatization_2021} use bibliographic data to measure the flow of researchers from Academia to industry and examine the factors driving it. They find that 25\% of the AI researchers moved to industry from institutions at the top 5 of the Nature Index. This observation suggests that industry tends to attract AI researchers from elite institutions, possibly reflecting a search for current and potential superstar talent or a narrow focus on high-prestige sources of talent. 

\citet{zhangEffectivenessAnalysisAltmetrics2019} employs altmetrics to assess the influence of AI publications, revealing an increased public interest in AI research findings since 2011. However, the literature suggests that altmetrics may not be suited to measuring societal impact \citep{bornmannAltmetricsPointBroader2014}.

The recent article of \citet{farberAnalyzingImpactCompanies2023} strikes to measure the relationship between authors' affiliations with the private sector and the article's popularity, measured by citations and attention score. They perform the keyword analysis with respect to the extent of association with the private sector. Their quantitative analysis shows the domination of the private sector in the AI research domain. 

\cite{kriegerAreFirmsWithdrawing2021} note an upward trend in private sector publications in fundamental research journals as opposed to those centered on applied research. They also highlight a rise in collaborative publications with academic entities in contrast to solo publishing efforts.

% content of papers: topic modeling, field of studies, values in papers
Several studies have tried to analyze the content of AI papers to study the influence of Big Tech. These include topic modeling of the abstracts. 
\citet{klinger_deep_2018} combine data from arXiv, GRID (Global Research Identifier) and MAG (Microsoft Academic Graph) to create a geocoded dataset of research activity in computer science disciplines. They identify deep learning papers using topic modeling. Finally, the authors measure the relatedness of computer science subjects based on their co-occurrence in arXiv papers.
\citet{klinger_narrowing_2020} analyze the field of study assigned to the papers based on arXiv. They find that companies focus more on applications of deep learning and on research advancing the computational infrastructure. AI techniques outside deep learning and broader AI applications, are of less interest to the private sector.   
\citet{birhane_values_2021} perform textual analysis in order to extract values encoded in papers. They analyze affiliations and funding sources in papers and find that the presence of Big Tech is increasing.

% Limitation of previous studies.
However, there is a significant gap in our understanding of Big Tech's influence. All of the above studies have at least one of the following limitations.

%Domination in citations
Firstly, none of the studies described above link the content of papers to citations in any way. Therefore, the resulting purely fraction-based analysis of content does not measure the spread of the ideas, but merely their prevalence. Quantifying both the spread and prevalence of specific ideas by using \textit{meme score} allows us to find the most contagious ideas, called \textit{memes}. The impact of affiliation groups on the contagiousness of each meme could be then modeled, resulting in a more fine-grained view of the influence of Big~Tech, Academia, and other groups. Moreover, this approach allows for a comparison of the ideas on which contagiousness is most affected by Big~Tech and Academia, which could provide insight into areas where Big~Tech or Academia has the most influence. 

% contagiousness of ideas
Secondly, they simplify the notion of influence to the share of affiliations in papers regarding a specific topic. They use the share of papers affiliated with a particular group only (with the exception of \citet{farberAnalyzingImpactCompanies2023}) that get published at prestigious conferences and in top journals as an indicator of the prevalence of Big Tech in AI research and interpret it as a proxy for influence measure. However, this approach does not take into account the number of the paper's citations, which is a more reliable proxy for the popularity of the paper and thus closer to measuring actual influence on the research field.

%Sources and size of the data
The third limitation is the quantity and diversity of data. The use of manual annotation limits the scope of the captured categories. Similarly,~using only arXiv or specific conferences as the source of the papers limits the representativeness of the data.

\subsubsection{Operationalization of the concept of Big Tech papers}
Author affiliation is a key characteristic associated with the author rather than the research paper itself, although even a simple quantitative approach (i.e., the number of authors) has been proven to be connected to the impact exerted by an article \citep{Sienkiewicz2016}. When examining the impact of Big Tech companies on scientific research, it becomes necessary to determine how to classify papers that are not exclusively authored by individuals affiliated with Academia or Big Tech. Existing approaches \citep{klinger_narrowing_2020,ahmed_-democratization_2020,birhane_values_2021} typically assume that any paper with at least one author affiliated with a Big Tech company should be categorized as a Big Tech paper. Alternatively, they consider the affiliation of the first author as the defining criterion.

However, these approaches oversimplify a more complex situation. Grouping papers authored exclusively by researchers from the private sector, along with papers primarily written by academics, one of whom may have a dual affiliation, seems counter-intuitive. It has been observed that researchers associated with Big Tech companies benefit from access to computational resources, tools, and datasets that would otherwise be unavailable to them \citep{whittaker_steep_2021,jurowetzki_privatization_2021}. Considering only the first author's affiliation would lead to overlooking these phenomena. 

Another approach is to look separately at papers that result from collaborations between Big Tech companies and other institutions. Articles~describing such collaborations tend to have different characteristics in terms of format and tone compared to the research papers mentioned above, and focus on the positive aspects of such collaborations \citep{popkin_how_2019}. In recent work, \citep{farberAnalyzingImpactCompanies2023} separately analyzed papers co-authored by Academia and company-affiliated researchers. To the best of our knowledge, no analyses have specifically examined the influence and topics of interest reflected in papers resulting from collaborations between Big Tech and academia.

Furthermore, we are not aware of any analysis that compares Big~Tech-affiliated papers to those affiliated with the private sector in general. This~limits the ability to assess whether found characteristics are specific to Big~Tech, or are phenomena visible in all research originating from the private sector.
% Difference between mixed and "pure" affiliation, the difference between Big Tech and companies in general
In our study, we examine the differences between several affiliation groups: purely Big Tech, purely academic, private sector as a whole, and papers affiliated with both Big Tech/private sector and Academia.

\subsection{Memetic analysis}
Several measures can characterize the spread of ideas in science \citep{Wagner2014,Xu2018} or in business research \citep{Wu2017}. A straightforward method to identify ideas within a set of documents is to examine their frequency. Techniques like tf-idf (term frequency–inverse document frequency) allow the discovery of potential keywords. Topic modeling methods, such as Latent Dirichlet Allocation (LDA) or BERTopic \citep{grootendorst2022bertopic}, help to understand the distribution of ideas consisting of multiple phrases across a document set. Another option is to focus on topical clusters, which can be built from bibliographic coupling and co-citation networks \citep{liu_knowledge_2017, liu_predicting_2019}. One can use a citations cascade \citep{min_citation_2021} or chains of citations \citep{della_briotta_parolo_tracking_2020}, which reveal a higher-order (i.e., reaching further than just the neighborhood of the node) temporal relationship between the papers.

% However, these methods do not provide insight into the dynamics of idea evolution or the flow of knowledge and concepts. Evolutionary analysis can help address such questions. 
One perspective on the ideas' propagation is looking at it through the lens of the evolutionary theory of science \citep{kantorovich_evolutionary_2014}. Central to this theory is the concept of \textit{meme}, first introduced in a broader context by \citet{dawkins_selfish_1976}. A meme is considered a cultural analog to a biological gene, a unit of information that replicates itself, mutates, and undergoes selection in the evolutionary process. Analogous to genes in biology, memes are pieces of information transmitted between individuals within a given culture.  

\citet{kuhn_inheritance_2014} proposed an operationalization of the concept of a~meme in the context of scientific knowledge by introducing the notion of a~\textit{meme score}. The meme score measures how much a given term is a meme by analyzing the citation network. It takes into account the term's frequency, the probability of its transmission through citation, and the probability of it emerging independently. In this operationalization of memes, a term exhibits more memetic quality, the higher the meme score of that term is. Since its proposal, the meme score has been used in several studies. It has been used to investigate how the gender of researchers influences their positions in the scientific community \citep{gender_authorship}, to identify scientific and technological trajectories of ideas in paper and patent networks \citep{Sun_Ding_2018}, and to explore diffusion cascades of ideas \citep{Mao_Liang_Cao_Li_2020}.

\section{Materials and methods}

\subsection{Meme score}

The \textit{meme score} was introduced by~\citet{kuhn_inheritance_2014} as a measure of the 'contagiousness' of meme.\par
The meme score for the given meme is calculated by multiplying the relative frequency of a term by its \textit{sticking factor} and then dividing it by its \textit{sparking factor}. The quotient of the sticking factor and the sparking factor is labeled as a propagation score. Both measures are briefly described below.

The sticking factor can be interpreted as the probability of the meme $m$~being transmitted from one paper to the other paper. It corresponds to the probability that the meme appears in the abstract of the given paper, given that it appears in at least one of the abstracts of the papers cited by it. It is calculated by dividing the number of papers that contain the meme $m$ and cite at least one paper with this meme by the number of papers that cite at least one paper with this meme: $d_{m\rightarrow m}$ by $d_{\rightarrow m}$. \par

The sparking factor can be interpreted as the probability of the meme appearing without being present in any of the cited papers - how likely it is that the meme will be mentioned in the abstract, given that it does not appear in any of the cited papers' abstracts. It is calculated by dividing the number of papers with the given meme that do not cite any papers with this meme $d_{m\rightarrow \cancel{m}}$ by $d_{\rightarrow \cancel{m}}$ all papers that do not cite any papers with the given meme. 

The formula for the meme score can be written as follows:

\begin{equation}
M_m = \frac{N_m}{N}\frac{d_{m \rightarrow m}}{d_{\rightarrow m}}\backslash \frac{d_{m \rightarrow \cancel m}}{d_{\rightarrow \cancel m}},
\label{eq:ms}
\end{equation}

where $N_m$ is the number of papers with the meme $m$, $N$ is the number of all memes in the dataset. $d_{m \rightarrow m}$ is the number of papers that contain the meme and cite at least one paper with the meme and $d_{ \rightarrow m}$ is the number of papers that cite at least one paper with the meme. On the other hand $d_{m \rightarrow \cancel m}$ is the number of papers containing the meme but not citing a paper with the meme, and $d_{\rightarrow \cancel m}$ is the number of papers that do not cite a paper with the given meme.

From the three components of the meme score, it is mainly the sticking factor that measures the meme's contagiousness. The other factors control if the phrase is not merely used as a part of a natural language or if a phrase is popular enough. Due to this, the sticking factor on its own could serve as a weaker measure of contagiousness. Meme score can be simply interpreted as a ratio of the probability of replicating a phrase to the probability of its appearance without earlier presence in the cited papers. To select meaningful memes, we must choose a threshold of the meme score defining which phrases are considered memes and which are not. This is done by selecting the point of maximum curvature from the value of the meme score as a function of its position in the phrase ranking ordered by the meme score (see Sec. \ref{res:memes} for practical implementation of the referred method). Such an approach is similar to the so-called ``elbow method'' \citep{Ketchen1996} -- a commonly used visual technique in assessing the optimal number of clusters in the k-means method.

\subsection{Conditioned sticking factor}

To measure how probable a meme is to spread under a given affiliation, we introduce a \textit{conditioned sticking factor} -- a modified version Kuhn's meme score. A similar approach was proposed in \cite{gender_authorship}, where the authors conditioned the meme score on the main author's gender. In contrast, we condition only the sticking factor on the authors' affiliations. The conditioned sticking factor is calculated as 
\begin{equation}
    \sigma_{m,a} = \frac{d_{m\rightarrow m,a}}{d_{\rightarrow m,a}}.
\end{equation}
The conditioned sticking factor divides the number of papers $d_{m\rightarrow m,a}$ that have meme $m$ and cite at least one paper with this meme and given affiliation $a$ by the number of papers $d_{\rightarrow m,a}$ that cite at least one paper with given meme $m$ and affiliation $a$. It can be interpreted as the probability of the meme transitioning during citation under the condition that the meme appeared in the paper with a given affiliation.\par

\begin{figure}
    \centering
    \includegraphics[width=0.66\textwidth]{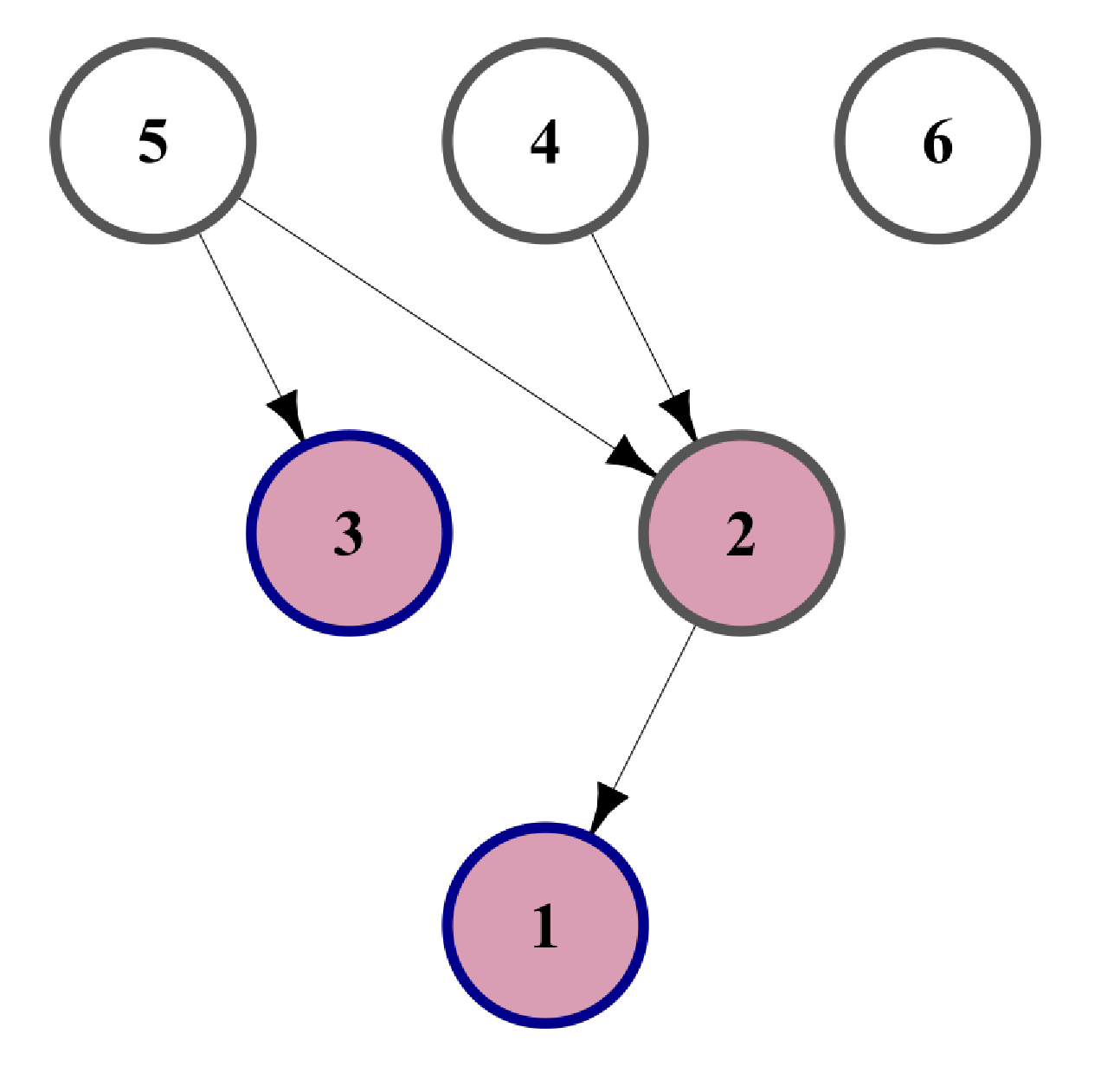}
    \caption{\textbf{Toy example of how to calculate the meme score.} Papers 1, 2 and 3~contain the meme (maroon background). Papers 1 and 3 have a given affiliation (dark blue frame) for which we calculate the conditioned sticking factor. The frequency of the meme is $\frac{3}{6}$. The unconditioned sticking factor is $\frac{1}{3}$, because of three papers that cite papers with the meme (2, 4, and 5), only one has the meme. The sparking factor is $\frac{2}{3}$ since the meme appears in two of three papers that do not cite papers with this meme – it appears in 1 and 3, but not in 6. The conditioned sticking factor is $\frac{1}{2}$ since the meme appears in 1~out of 2 citations of the paper with the meme and given affiliation (2 replicates the meme from 1, 5 does not replicate the meme from 3). }
    \label{fig:toyexample}
\end{figure}

In contrast to the earlier approach \citep{gender_authorship}, we have decided not to condition the sparking factor. The original purpose of the sparking factor is to control how often the word appears without any influence and limit the appearance of the most popular words that do not convey specific ideas. The sparking factor, conditioned analogously to the sticking factor, would correspond to the probability that the meme appears in a paper without being present in any cited papers with a given affiliation. However, the meme can appear in the other cited papers with different affiliations. In this way, the conditioned sparking factor would lose its original purpose of excluding phrases not conveying ideas without providing us with any specific additional information.
For this reason, we decided to condition only the sticking factor. Finally, once $N_m \gg d_{m \rightarrow m}$, i.e., the meme is relatively common but rarely transmitted through the citation network, the meme score can be approximated by the sticking factor. Due to this, we focus primarily on the relationships between conditioned sticking factors for different groups. Figure \ref{fig:toyexample} presents a toy example to explain how the sticking factor, sparking factor, meme score, and conditioned sticking factor are calculated. 

\subsection{Papers dataset and processing}

As a main data source for this study, we used the S2ORC database \cite{Lo2020S2ORCTS},~which is a large corpus of 81.1 million English-language academic papers spanning many disciplines. The collection comes from various sources, such as in-proceedings of scientific conferences, well-established journals, and digital archives (e.g. arXiv). We gathered 557~681 articles from this source by searching for phrases presented in  \ref{appendix:keywordslist}, which occurred in the titles and abstracts of the papers. The aforementioned set of keywords is based on the list proposed by \cite{liu_tracking_2021}, who conducted the study, attempting to find a representative set of keywords for the AI domain while maintaining a reasonable balance between recall and precision (we consider possible data limitations in the Discussion section). 

Afterwards, we performed exploratory data analysis. We found out that 138~878 records had no DOI identifiers, which we used to match the papers against the OpenAlex database. As affiliation data provided by those links is crucial for our study, we decided to remove such papers from our dataset. Next, we discovered that 146~525  of the remaining records had no information about citations. Papers like this were useless for citation network analysis, so we decided to remove them from our dataset. Finally, we ended up with 166~455 records that had affiliations matched with OpenAlex.

The whole pipeline of memes extraction is shown in Fig.~\ref{fig:data-proc-pipeline}. All used abstracts were preprocessed first -- we lowercased all words and tokenized them. For the extraction of noun chunks, we used the \texttt{spaCy} library with the \texttt{en\_core\_web\_md} model.

\subsubsection*{Papers' affiliation}
First, we look at the category of the company as it is specified in the OpenAlex~Database. OpenAlex provides information about the type of institution -- it categorizes them into the following categories: company, education, government, facility, healthcare, nonprofit, and other (Table~\ref{tab:annotations}). By analyzing these categories, we decided to merge education, facility, and government into Academia. We also analyzed the company category separately.\par
As Big Tech, we take a subset of technological companies that were among the top 60 companies by market capitalization in the years 2016-2019 \citep{pwc2019,pwc2018,pwc2017,pwc2016}. The following companies fall into this category: Apple, Google, Microsoft, Facebook, Tencent, Oracle, Intel, IBM, Cisco Systems, TSMC, SAP, Qualcomm, Amazon, Siemens, Alibaba Group, Nvidia, and Samsung. We aimed to construct a list that encompasses companies previously labelled as Big Techs in media, such as the Big Five, FAANG or MAGA \citep{big5economist}, while also being supported by a quantitative measure like market capitalization. To ensure the list's robustness and account for short-term market changes, we use a union of companies that appeared in the top 60 companies during the four years with the highest number of papers in our dataset. We opt to limit ourselves to the top 60 companies to include companies considered most significant to the field while excluding the lower end of the top 100 companies, where most technological companies appeared only once during the four years we consider. Papers affiliated with Big Tech are 38\% of the papers affiliated with the company category, but they account for 57\% of their citations. \par

\section{Results}

\subsection{Non-binary distinction between Big Tech and academy--affiliated papers}
\label{sec:dist}

\begin{figure}[!ht]
    \centering
    \includegraphics[width=.7\textwidth]{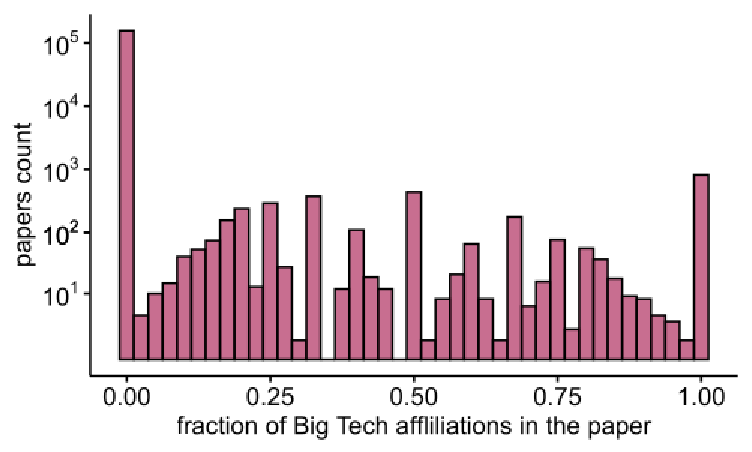}
    \caption{\textbf{Histogram (log scale) of the number of papers for different fractions of Big Tech affiliations}. The length of each bin is 0.025. Most papers are affiliated only with Academia, yet we can also witness a visible peak for purely Big Tech papers, which underlines the validity of this group. The plot also suggests that we should consider a~third group of mixed affiliations, as the number of articles with Big Tech fraction is even bigger than that of Big Tech papers.}
    \label{fig:aff_histogram}
\end{figure}

We provide substantial arguments, advocating for the less coarse distinction between Academia and Big Tech affiliations than the bisection often used in the literature. The most natural way to follow this proposition would be to use a continuous space by simply calculating for each paper the proportion of Big Tech affiliations to the total number of affiliations. However, as shown in Fig.~\ref{fig:aff_histogram}, these values are unevenly distributed  -- the vast majority are simply academic papers (almost 98\% of all papers in the examined dataset), around 0.5\% belong only to the Big Tech class, and the remaining part is characterized by a mixed affiliation. The last group also suffers from obvious finite-size effects, i.e., due to a limited number of authors,  fractions such as $\frac{1}{4}, \frac{1}{3}, \frac{1}{2}, \frac{2}{3}$, and $\frac{3}{4}$ tend to occur more frequently than other values (see Fig.~\ref{fig:aff_histogram}). To avoid the above obstacles and to be consistent with the observations presented in Fig.~\ref{fig:aff_histogram}, we decided to use a \textit{ternary} affiliation scheme, i.e., Academia--mixed Big Tech--Big Tech.

\begin{figure}
    \centering
    \includegraphics[width=\textwidth]{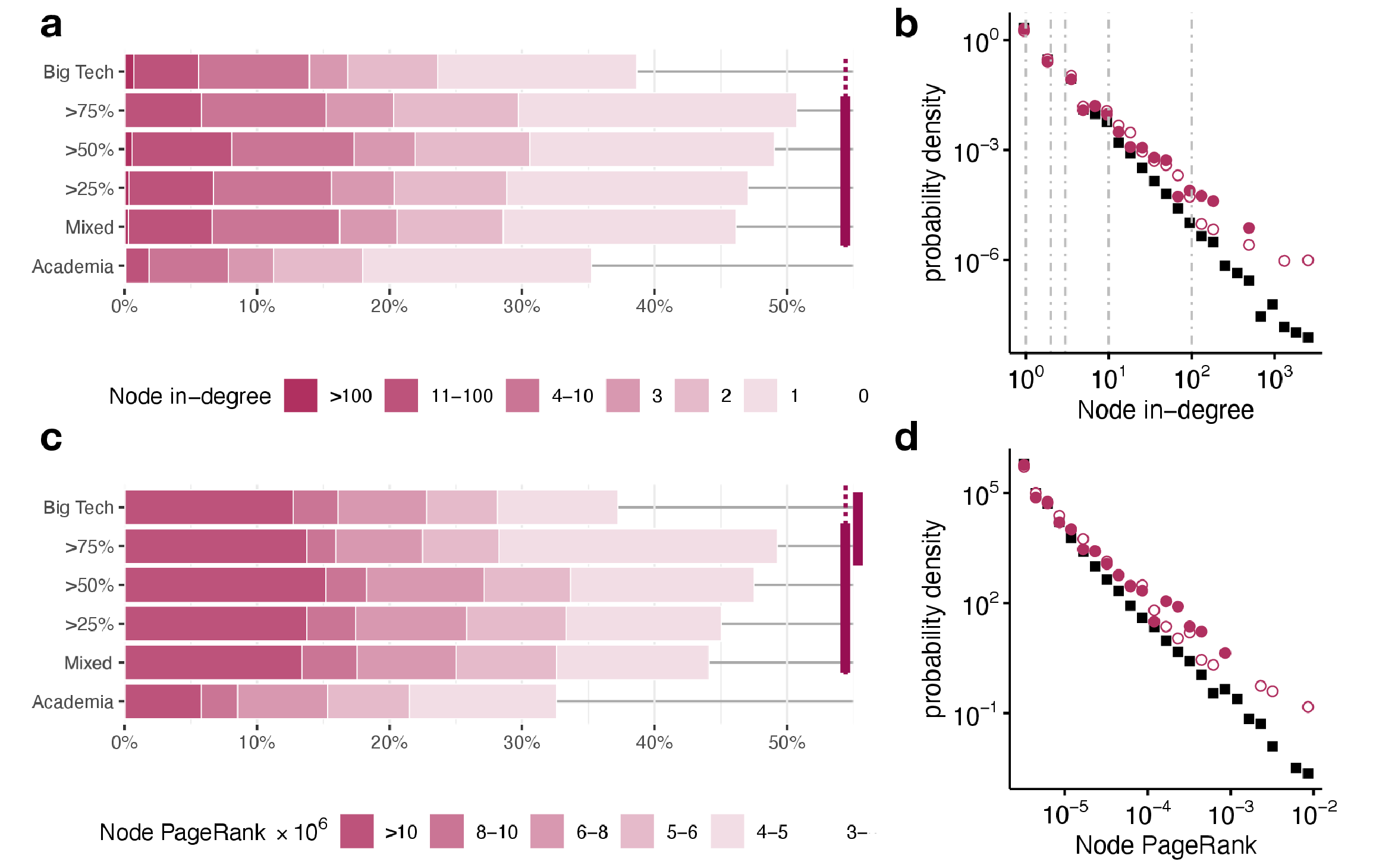}    
    \caption{\textbf{Differences between Academia--mixed Big Tech--Big Tech as seen from the citation network perspective}. The rows are, respectively nodes' in-degrees (panels a, b) and the PageRank values (panels c, d). The left column shows the distribution plot, presenting the percentage of nodes with a given in-degree or PageRank value for ternary classification (``Academia'', ``Mixed'', and ``Big Tech``) as well as cumulative quartiles (see text for details). Vertical solid lines on the far right of panels a and c connect statistically indistinguishable categories, while vertical dotted lines extend solid lines for the case of the network with removed isolated nodes. The right column presents probability density distributions associated with the ternary classification on panels a and b: filled squares represent Academia, empty circles -- mixed Big Tech, and filled circles -- Big Tech (node in-degree is increased by one to overcome log scale issues).}
    \label{fig:network_analysis_Big Tech}
\end{figure}

Let us first explore the implications of the ternary division using simple network metrics: incoming degree and PageRank \citep{Brin1998}. Both these measures can be used to quantify the importance of a node (paper). In the case of the in-degree, we simply tally the number of citations a paper has received, while PageRank provides more nuanced information. Figure~\ref{fig:network_analysis_Big Tech} shows in-degree (panel a) and PageRank (panel c) distributions, grouped according to the introduced ternary affiliation classification (denoted as ``Academia'', ``Mixed'' and ``Big Tech''). In both cases, the Kruskal-Wallis test finds significant differences among all groups (p-value $< .001$), while post-hoc pairwise comparisons using Dunn's test with Holm-Bonferroni corrections indicate that the described distinctions are maintained for all pairs of groups (if two or more categories are statistically indistinguishable, they are connected with a solid line on the far right part of the panel). To prove that the division into three categories does not lead to a loss of information available in the continuous approach, we divide the dataset into six categories: the first two and the last one are the same as in the previous division, while the remaining three describe cumulative quartiles without exclusive Big Tech affiliations (``$>25\%$'', ``$>50\%$'' and ``$>75\%$'', i.e., ``$>25\%$'' means that each node in this category has over 1/4 share of Big Tech affiliations but cannot be exclusively Big Tech). We obtain similar results for the Kruskal-Wallis test (significant differences with p-value $< .001$) as well as for pairwise comparisons, the only exception being the penultimate category (``$>75\%$''), which is indistinguishable from the pure Big Tech affiliations in the case of PageRank. All the intermediate categories (``Mixed'', ``$>25\%$'', ``$>50\%$'' and ``$>75\%$'') are indistinguishable from each other, proving that further divisions in the ``Mixed'' category lack additional insights.

To examine the reasons for the clear differences between the elements of our ternary classification, we focus on the in-degree and PageRank probability distributions shown in Fig.~\ref{fig:network_analysis_Big Tech}b and Fig.~\ref{fig:network_analysis_Big Tech}d, respectively. Clearly,~mixed and Big Tech groups differ from Academia by higher probability density values for larger values of in-degree or PageRank. On the other hand, the difference between mixed and Big Tech is not as pronounced. The likely reason for the observed differences is connected to the number of zero-degree nodes -- the proportion of such nodes in Academia, mixed and Big Tech is 0.648, 0.538, and 0.613, respectively. Although the ratio of zero-degree nodes in the mixed category is visibly lower than in the other two categories, we confirm this observation by performing pairwise two-sample proportion tests with Holm-Bonferroni adjustment, which give p-values of $p < .0001$ for Academia--mixed Big Tech, $p=.044$ for Academia -- Big Tech, and $p = .0004$ for mixed Big~Tech--~Big~Tech. Thus, the network with removed zero-degree nodes brings different statistical results marked in Fig.~\ref{fig:network_analysis_Big Tech}a and Fig.~\ref{fig:network_analysis_Big Tech}c by the dotted line on the far right -- in this case, we recover the binary division as only ``Academia'' category is distinguishable from any other. 

\begin{figure}
    \centering
    \includegraphics[width=\textwidth]{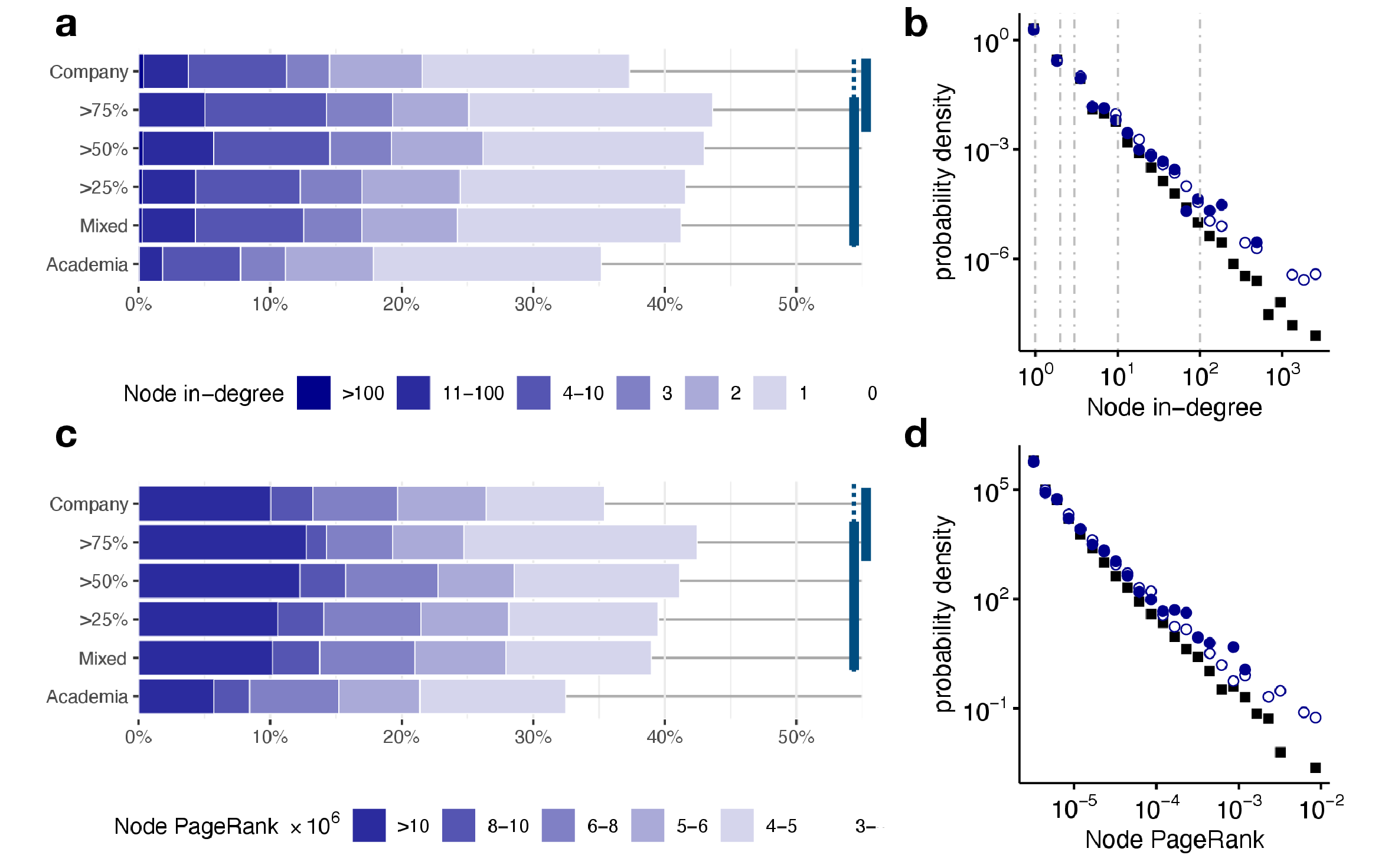}

    \caption{\textbf{Differences between Academia--mixed Company--Company as seen from the citation network perspective}. The rows are, respectively, nodes' in-degrees (panels a, b) and the PageRank values (panels c, d). The left column shows the distribution plot, presenting the percentage of nodes with a given in-degree or PageRank value for ternary classification (``Academia'', ``Mixed'', and ``Company``) as well as cumulative quartiles (see text for details). Vertical solid lines on the far right of panels a and c connect statistically indistinguishable categories, while vertical dotted lines extend solid lines for the case of the network with removed isolated nodes. The right column presents probability density distributions associated with the ternary classification on panels a and b: filled squares represent Academia, empty circles -- mixed Company, and filled circles -- Company (node in-degree is increased by one to overcome log scale issues).}
    \label{fig:network_analysis_cmp}
\end{figure}

Lastly, the decision to introduce the Big Tech category instead of using OpenAlex company category was supported by the intuition that Big~Techs differ significantly from other company actors and that extrapolating conclusions from all private companies to Big Tech would be unjustified. To test whether Big Techs are different from the rest of the private companies, we compare the PageRanks of Big Tech–affiliated and company-affiliated papers, performing the Kruskal-Wallis test between the PageRanks of Big Tech--affiliated papers and papers affiliated with companies that are not Big Tech. The results confirm that the distribution of PageRanks of Big Tech-affiliated papers is indeed different from that of other companies (p-value < 0.05). Nevertheless, analyzing the ternary division from the company's point of view (instead of Big Tech) does not bring any additional insights, as can be seen in Fig.~\ref{fig:network_analysis_cmp}. 

\subsection{Most contagious ideas in AI research}\label{res:memes} 

Using the meme score measure, we identified and analyzed which ideas are the most prevalent in AI research in general, which is the first step in differentiating Big Tech from Academia. To perform this analysis, we calculated the meme scores and propagation scores and counted the occurrences for all memes in our dataset, resulting in over 1.7 million observations. Such an amount of data had to be filtered to get the most interesting results. Thus, we decided to limit our study to phrases where the meme score was greater than 0, resulting in over 60,000 meaningful observations.

\begin{figure}
    \centering
    \includegraphics[width=\textwidth]{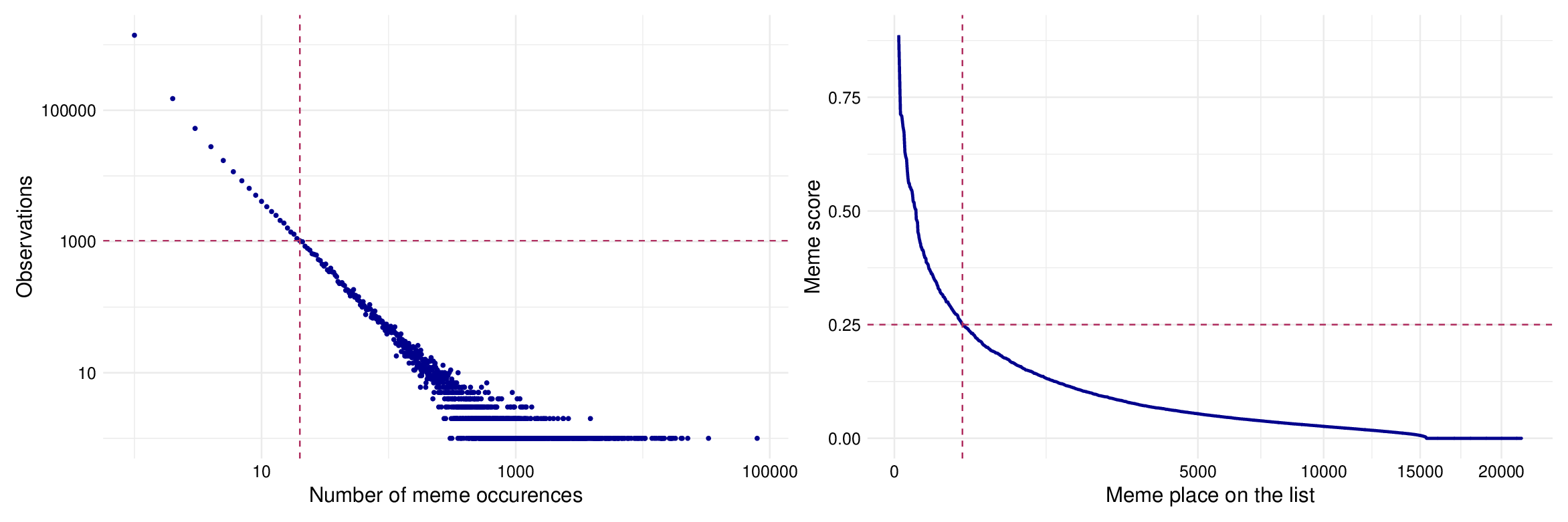}
    \caption{\textbf{Top memes selection by thresholds.} The left plot shows the number of remaining phrases, depending on the minimum number of meme occurrences. Since the dispersion of observations occurs around the number 20 on the x-axis, and 1000 on the y-axis (marked as dashed lines), we decided to use this point as a cutoff, which resulted in limiting ourselves to over 20 000 observations that have more than 20 appearances. The right plot shows the meme score value of the observation as a function of its position in the list sorted by the meme score. Since the major flattening occurs around the meme score value of 0.25, we decided to use this point as a cutoff, which resulted in limiting ourselves to 251 observations (marked as dashed lines).}
    \label{fig:meme_occurences_and_meme_score_thresholds}
\end{figure}

Since our main goal is to analyze the most spreading ideas deeply, we limited the number of memes according to two criteria: the observed occurrences and their meme score values. The first criterion is based on the idea that the memes with high meme scores that only occur once or twice are not meaningful topics, as they are too niche to be considered contagious. The second condition is obvious since the higher the meme score, the more likely it is to spread. We analyzed the curves shown in Fig. \ref{fig:meme_occurences_and_meme_score_thresholds} to choose a cutoff value. Finally, after limiting ourselves to phrases with a meme score above 0.25 and at least 20 occurrences, we ended up with 251 observations representing the top memes in our dataset. A manual annotation and inspection of the selected memes yields 80\% precision, which is a comparable result to the one reported by \citealt{kuhn_inheritance_2014} (i.e., around 81.2\% on 150 memes). Let us underline that it is possible to exclude obviously meaningless memes like `paper' or `results,' however, it requires introducing a free parameter into Eq. (\ref{eq:ms}), in a similar way as in \cite{kuhn_inheritance_2014}. Another option is a manual intervention. We refrained from following either of these paths to maintain high clarity of methodology.  

The set of chosen top memes was later annotated by the group of domain experts, which is briefly described in \ref{appendix:Annotations}, resulting in the assignment of one category per observation. The resulting categories are \textit{ML algorithms}, \textit{Medical terms}, \textit{Security}, \textit{Niche ML applications}, \textit{ML Concept}, \textit{Maths}, \textit{Graphs}, \textit{Computer Vision}, \textit{Natural Language Processing}, \textit{Data Related}, and \textit{Other}. The annotations of the top memes allow us to analyze which high-level topics are popular among ML researchers, indicating the most promising research areas. 
According to Table \ref{Memes_summary}, the most popular topic among scientists is the medical applications of ML. This category has the most memes and a third number of the memes occurrences in our dataset. It is not the only application, however, as we can see on the top memes list other important topics like security, and a whole category describing niche applications of ML solutions. However, this does not mean that today's research focuses only on ML applications, as the second-highest average meme score value, and the second sum of the occurrences belong to the topic describing the ML algorithms. Additionally, we were able to distinguish some particularly popular directions of AI development, such as Computer Vision or NLP. 
Finally, such an in-depth analysis proves the quality of our framework as the category Other, which includes unrelated topics, as well as meaningless memes, is responsible for only 20\% of detected memes. The group has the biggest number of occurrences due to the memes such as `paper', `results', or `art' (part of `state of the art' phrase), which occur multiple times in the majority of the papers.    

\begin{table}[!ht]
    \centering\renewcommand\cellalign{c}
    \setcellgapes{3pt}\makegapedcells
    \begin{tabular}{lcccc}
    \hline
        Category & \makecell{Number of \\ memes} & \makecell{Mean \\ meme score} & \makecell{Mean \\ occurrences} & \makecell{Sum of \\ occurrences} \\ \hline
        Medical term & 70 & 0.339 & 141 & 9865 \\ 
        Other & 50 & 0.347 & 2620 & 130993 \\ 
        Niche ML application & 24 & 0.345 & 56 & 1336 \\ 
        Security & 23 & 0.346 & 107 & 2454 \\ 
        ML Algorithm & 20 & 0.380 & 581 & 11616 \\ 
        ML Concept & 20 & 0.343 & 73 & 1456 \\ 
        Computer Vision & 17 & 0.331 & 79 & 1336 \\ 
        Maths & 9 & 0.348 & 35 & 312 \\ 
        Data Related & 7 & 0.405 & 48 & 338 \\ 
        NLP & 7 & 0.332 & 204 & 1430 \\ 
        Graphs & 4 & 0.384 & 47 & 188 \\ \hline \\
    \end{tabular}
    \caption{\textbf{A statistical summary of the topics present in the top memes subset.} The table provides information about the number of memes existing within the described category. It also shows additional information describing the average meme score value, the average number of meme occurrences, and the sum of the total occurrences of memes from that topic.}
    \label{Memes_summary}
\end{table}

\begin{sidewaysfigure}
    \centering
    \includegraphics[width=19cm]{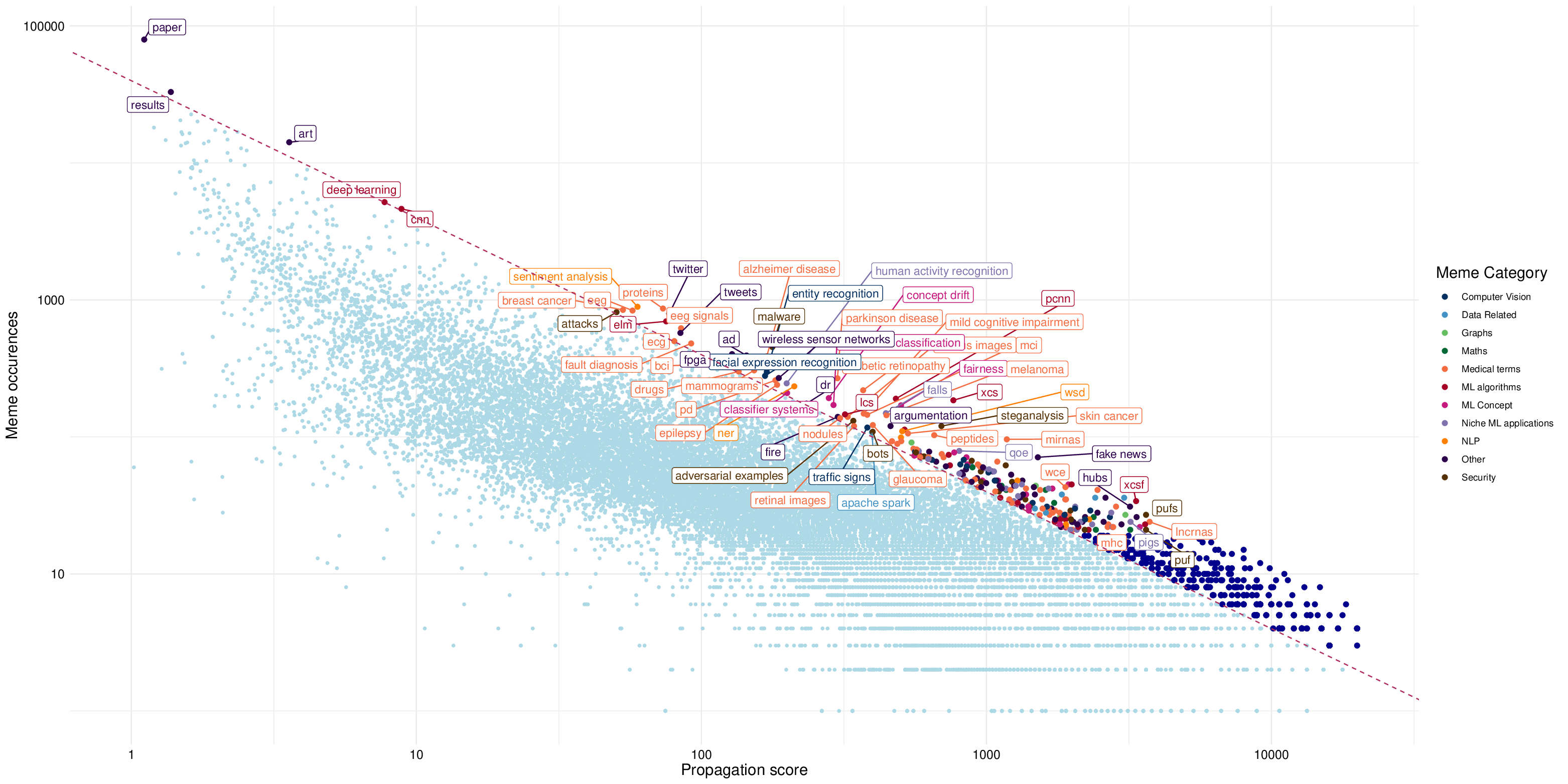}
    \caption{\textbf{Characteristics of top memes compared with other observations.} The observations on the plot are memes present in the whole dataset, and they are distinguished by the number of occurrences and propagation scores. Light-blue observations are memes with a meme score below 0.25, whereas the dark-blue ones have a score above 0.25, but the number of occurrences of these observations is below 20. The colorful dots represent the top memes with meme scores over 0.25, and more than 20 occurrences. These observations are additionally assigned to the meme categories described in the paragraph above. Exemplary observations were also labeled with their meme name.}
    \label{fig:saddleplot}
\end{sidewaysfigure}

Knowing the big picture of scientist's interests, we can combine it with a more detailed analysis, providing a wider perspective for our top memes group.

Figure \ref{fig:saddleplot} shows the characteristics of the top memes subset. As the meme score value depends on the number of meme occurrences and the propagation score, we can imagine that reflecting a meme score equal to 0.25 is used as a cutoff value in this analysis. From the density of observations, we can grasp the tendency for the highest meme scores to be assigned to the observations with fewer occurrences. Additionally, the plot helps us detect the outliers aforementioned in the previous section, as they yield low propagation score, but also have many occurrences. Thanks to the exemplary meme annotations, we can also track some memes and analyze the reasons why they are considered important. For example, `fairness' is considered important mostly due to its high propagation score, whereas `cnn' or `deep learning' benefits from a vast number of occurrences.

\begin{sidewaysfigure}
    \centering
    \includegraphics[width=19.5cm]{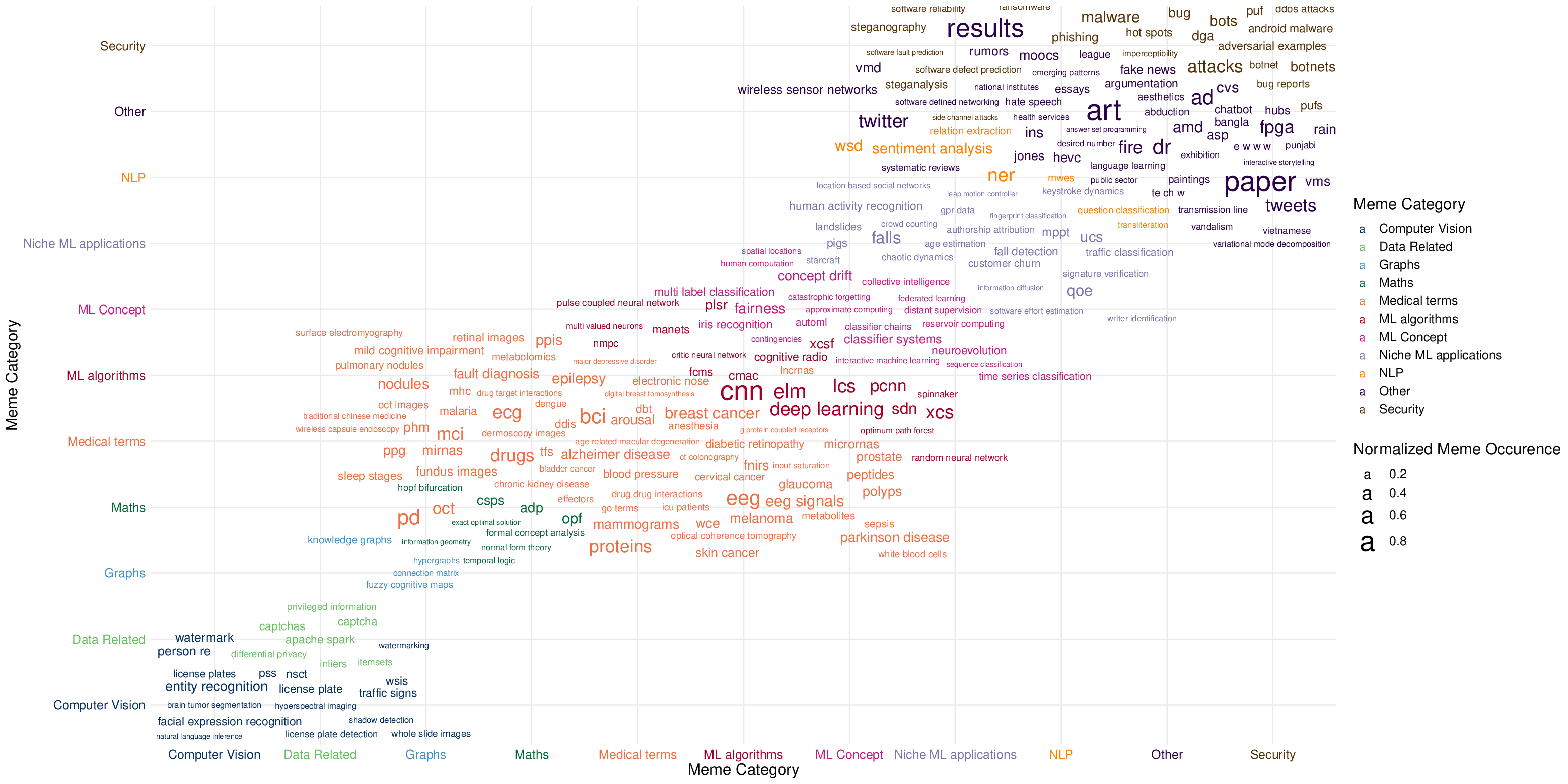}
    \caption{\textbf{A word cloud plot presenting top memes.} The size of the text represents the meme occurrence normalized to a [0, 1] range, whereas the color shows to which topic the meme belongs. The word cloud is organized in such a way that observations from the same topic are close to one another, as the X, and Y axes are discrete and consist of ordered categories.}
    \label{fig:wordcloud}
\end{sidewaysfigure}

To finalize the topic analysis, we present all top memes as a structured word cloud shown in Fig. \ref{fig:wordcloud}. This way, we can analyze particular memes that build each of the predefined categories. We should particularly focus on the medical memes, as this topic is definitely the most important one due to the abundance of different memes. The memes creating this topic (presented as orange) cover various medical terms. We~can discover which disease treatments and studies greatly benefit from the usage of ML (e.g., various cancers, epilepsy, malaria, etc.). We also can indicate that another major points of interest are the studies considering genomics (lncRNA, miRNAs, etc.), and that ML is widely used for various examinations (mammograms, eeg, ecg, etc.).

\subsection{Differences in contagiousness between Companies, Academia, and Big Tech}
\begin{figure}
    \centering
    \includegraphics[width=.85\linewidth]{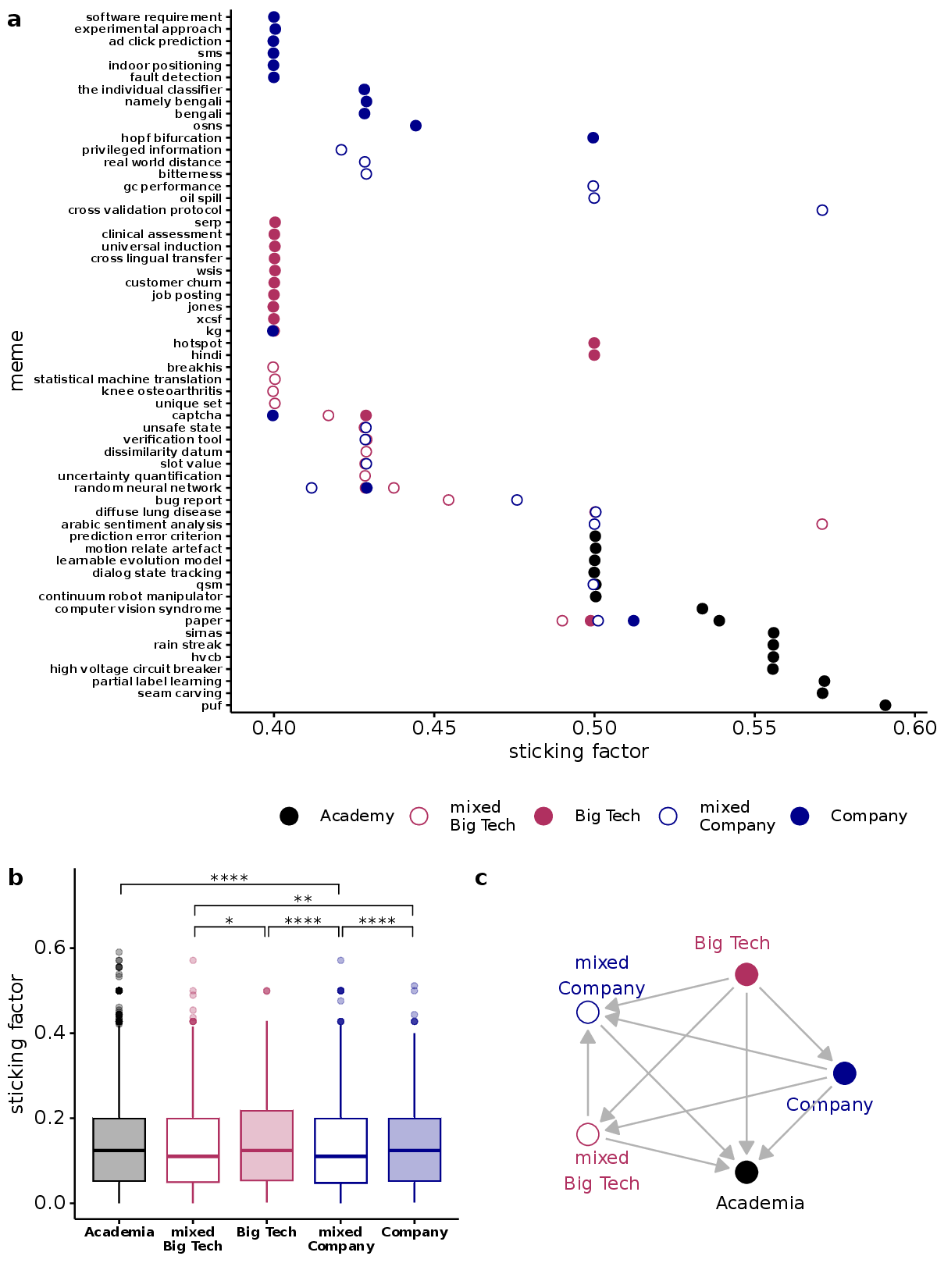}
    \caption{\textbf{Differences in contagiousness between Companies, Academia, and Big Tech.} (a) Selection of 15 memes with the highest sticking factor values in each category. (b) Boxplots of sticking factors conditioned on affiliation. The significance of post-hoc comparisons is indicated by *, **, ****, representing $.01 < p < .05$, $.001 < p < .01$ and $p < .0001$, respectively -- insignificant comparisons are not shown. (c) Results of Wilcoxon tests for paired memes -- an incoming arrow (pointing from category $X$ to category $Y$) informs about statistically higher sticking factors of paired memes in $X$ than in $Y$.}
    \label{fig:sticking-factor-dunn}
\end{figure}

To assess whether contagiousness varies depending on author affiliation, we conducted a comparative analysis of conditioned sticking factors across different affiliation groups. A selection of the 15 highest-ranked memes in each affiliation (Academy, mixed Big Tech, Big Tech, mixed Company, and Company) ordered according to their conditioned sticking factor is shown in Fig. \ref{fig:sticking-factor-dunn}a. Although this limited picture suggests that \textit{Academia} memes dominate, we need to underline that this is simply an illustrative excerpt from the whole distribution. 

In a more systematical analysis, we compared all non-zero conditioned sticking factors across all affiliation groups using Kruskal-Wallis test, indicating a significant difference across groups in terms of the sticking factor, i.e., affiliation is a factor that creates distinctions among sticking factor values.  Outcomes of post-hoc comparison (Dunn's test) between categories presented in Fig. \ref{fig:sticking-factor-dunn}b pinpoint distinctions among the sticking factor distributions: strikingly, there are no statistical differences among Academia, Big Tech, and Company. On the other hand, such pairs as mixed Big Tech -- Big Tech or mixed Company -- Company can be distinguished. This is opposite to the relations observed using network metrics for non-isolated nodes shown in Figs. \ref{fig:network_analysis_Big Tech} and \ref{fig:network_analysis_cmp}.    

The above analysis is conducted in absolute terms, i.e., the distributions can be biased because Academia affiliations are more frequent than Big Tech or Company ones (cf Fig. \ref{fig:aff_histogram}). To overcome it, we focused on memes that appear in at least two different affiliation categories, e.g., ``random neural network'' (present in mixed Company, Company, mixed Big Tech, and Big Tech) depicted in Fig. \ref{fig:sticking-factor-dunn}a. Following, we conducted Wilcoxon signed-rank tests, comparing sticking factors conditioned on specific groups using matched samples (the same memes in both groups). If we were able to reject the null hypothesis (observations in both groups $X$ and $Y$ are exchangeable) in favor of the one-sided alternative hypothesis (differences $X-Y$ are stochastically larger than a distribution symmetric about zero), we noted it with a link in Fig. \ref{fig:sticking-factor-dunn}c, pointing from $X$ to $Y$.  The results revealed that memes affiliated with Big Tech exhibited higher sticking factors compared to all other affiliation categories, with company-affiliated memes ranking second. On the other hand, academic affiliation was associated with a decrease in the sticking factor compared to other affiliations.

\begin{figure}
    \centering
    \includegraphics[width=\textwidth]{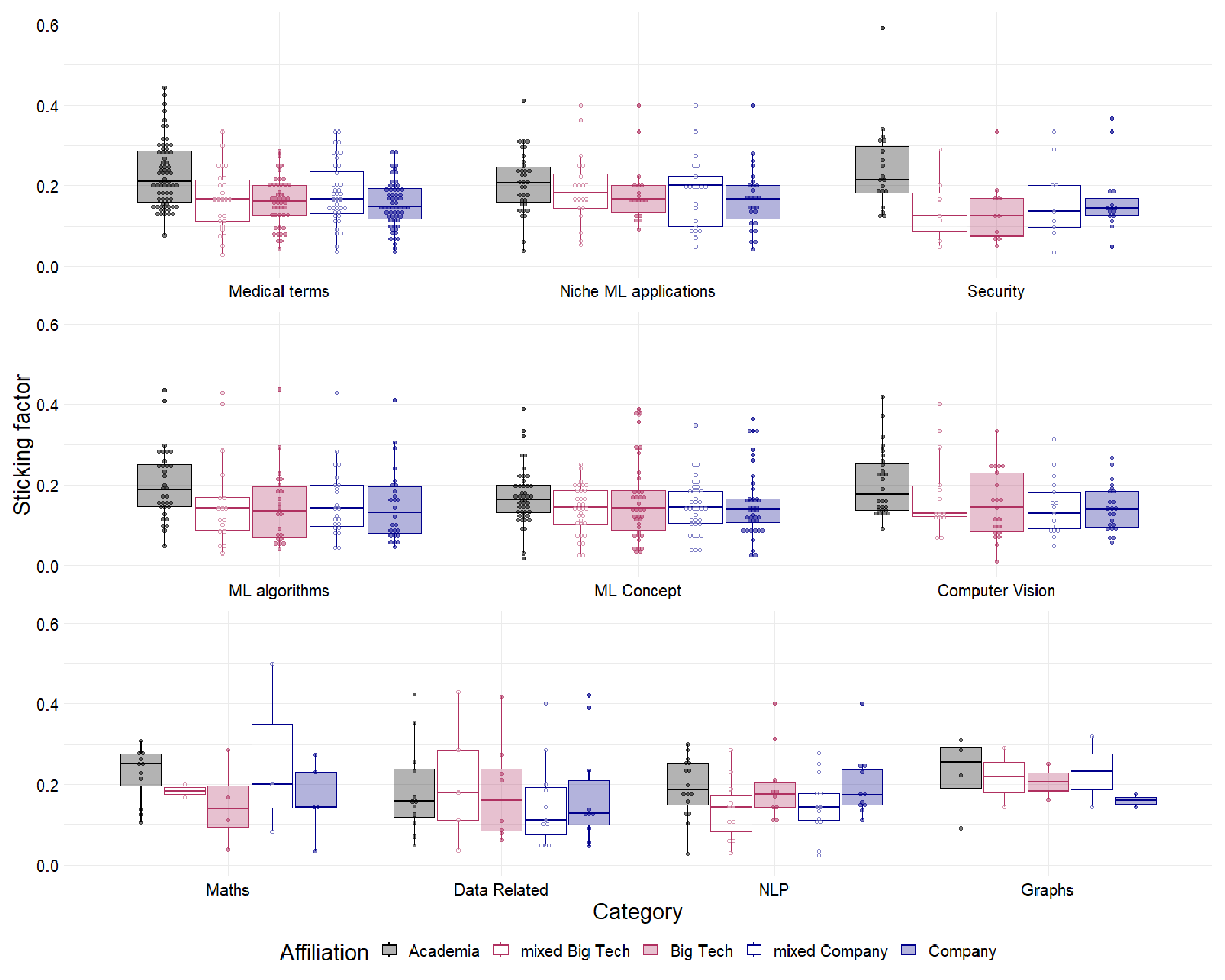}
    \caption{\textbf{Conditioned sticking factor across ML categories.} Box-plot depicts the distribution of the conditioned sticking factor of memes across different ML categories, stratified by the originating affiliation. The distribution patterns for most categories mirror the aggregate trends. In the `Data Related' segment, memes generated by Big Tech exhibit a noticeably higher sticking factor, suggesting a dominant role in propagating novel content within this domain—likely a reflection of the substantial resources they possess. For `Niche ML Applications', companies have a higher median sticking factor compared to their performance in other meme categories, implying their effectiveness in spreading memes that may complement their commercial strategies. }
    \label{fig:conditioned-sf-categories}
\end{figure}

We then explored how this difference depends on the meme category introduced in Table \ref{Memes_summary}. Five hundred memes with the highest meme scores, excluding those not present at least once in papers from each affiliation group, were selected for this comparison. After removing duplicates and the least common categories of memes, the resulting list has 148 memes. The comparison of conditioned sticking factors divided by meme category is presented in Fig. \ref{fig:conditioned-sf-categories}.

\section{Discussion and conclusions}\label{sec:diss}

This study aims to identify and measure the contagiousness of AI research ideas (memes) based on authors' affiliations and whether this differs between pure and mixed academic or company affiliations. We used two metrics, the meme score, and the sticking factor, and introduced the conditioned sticking factor to assess this relationship.

Our research reveals that papers co-authored by individuals from both Big Tech and academic institutions tend to receive significantly higher citation counts. Additionally, papers co-authored by individuals from both corporate and academic backgrounds exhibit distinct scientific characteristics compared to papers authored exclusively by academia or corporate entities. These findings challenge the binary distinctions often assumed in prior research.

The analysis suggests that while the contagiousness of memes authored by academia does not significantly differ from those by companies or Big Tech on average, memes mentioned by Big Tech have higher contagiousness when compared to academia. This difference in contagiousness could be linked to the diversity of memes produced and propagated by the two groups, with academia creating far more memes due to the difference in overall papers authored. The variance in contagiousness may also be attributed to factors such as the availability of resources for Big Tech's researchers.

The impact of affiliation groups on the spread of memes can be understood through framing theory. This theory suggests that different perspectives and values influence how issues are perceived \citep{chongFramingTheory2007}. In the context of memes, the affiliations of creators contribute to different framings of the same meme. The contagiousness of a meme within a specific group indicates that group's influence over the framing of the meme. Analyzing how memes co-occur and their connections to affiliations can provide further insight.

In our analysis, we explore the impact of Big Tech and academic institutions on the propagation of ideas in scientific papers. We introduce a new concept, the conditioned sticking factor, which can be used to analyze how various factors influence the spread of ideas. Our approach could help track relations between organizational characteristics and academia-industry interactions \citep[see, e.g.,][]{Scandura2022} and possible outcomes could complement or challenge the results obtained in previous studies that tackled the problem of diffusion of ideas in a given scientific discipline \citep{Hargreaves2005}, or technology adoption under different (also institutional) constraints \citep{Galang2014} Additionally, the conditioned meme score may aid in measuring information overload in science \citep{Holyst2024}, a critical issue for researchers.

Further direction of research should consider the dynamics of memes' contagiousness and investigate the impact of affiliation on contagiousness. Comparing the relationship between Big Tech and Academia in AI with similar relations in non-AI computer science and in other disciplines like medicine \citep{Giunta2016} would provide valuable insights. A limitation of this study is the operationalization of memes as simple noun chunks, overlooking the synonymity of phrases and, consequently, potentially high-spreading memes. Possible solutions could be techniques such as clustering embedded memes. Another potential constraint comes from the method of selecting the keywords to filter AI-related papers -- as mentioned in \ref{appendix:keywordslist}, we only partially used the approach suggested by \cite{liu_tracking_2021}, deeming another set of secondary and ternary keywords being suitable. One needs to mention that \cite{farberAnalyzingImpactCompanies2023}, who pursued a similar topic of academia and company affiliations in AI research, focused on the Microsoft Academic Graph field of study. In the same manner, \cite{gao_pasteurs_2024} relied on Web of Science classification, examining if patent-cited papers have a higher impact. All these examples show a lack of a common gold standard that might limit the generality of obtained conclusions. In recent years, non-profit research organizations like OpenAI have made notable contributions to the AI community, but their research output has not yet reached the scale of Big Tech (see Fig. \ref{fig:openai-citations} in \ref{appendix:openai} for specific numbers). These organizations, such as Anthropic, have innovative approaches to disseminating research findings using non-traditional publishing methods with interactive features. However, quantitatively analyzing their impact poses challenges, as their publications are excluded from standard paper indexing databases. Our analysis is also restricted to the year 2020, preventing the observation of memes related to generative AI applications.

Despite these limitations, our study suggests that the notion of Big Tech's dominance over AI research is overly simplistic, advocating for a more nuanced understanding of the roles played by Big Tech companies, corporations, and Academia in shaping the future of AI.
\section*{Acknowledgements} 
The research was funded by (POB Cybersecurity and Data Science) of Warsaw University of Technology within the Excellence Initiative: Research University (IDUB) programme. This work was also funded by the European Union under the Horizon Europe grant OMINO – Overcoming Multilevel INformation Overload (grant number 101086321, \url{https://ominoproject.eu}). Views and opinions expressed are those of the authors alone and do not necessarily reflect those of the European Union or the European Research Executive Agency. Neither the European Union nor the European Research Executive Agency can be held responsible for them. Computational part of this study was supported in part by the Poznań Supercomputing and Networking Center (grant number 607).

\bibliographystyle{plainnat}
\bibliography{bibliography}  %%% Remove comment to use the external .bib file (using 
\newpage

\appendix
\section{Data gathering and processing}
\label{appendix:keywordslist}

\begin{figure}[!hb]
    \centering
    \includegraphics[width=\textwidth]{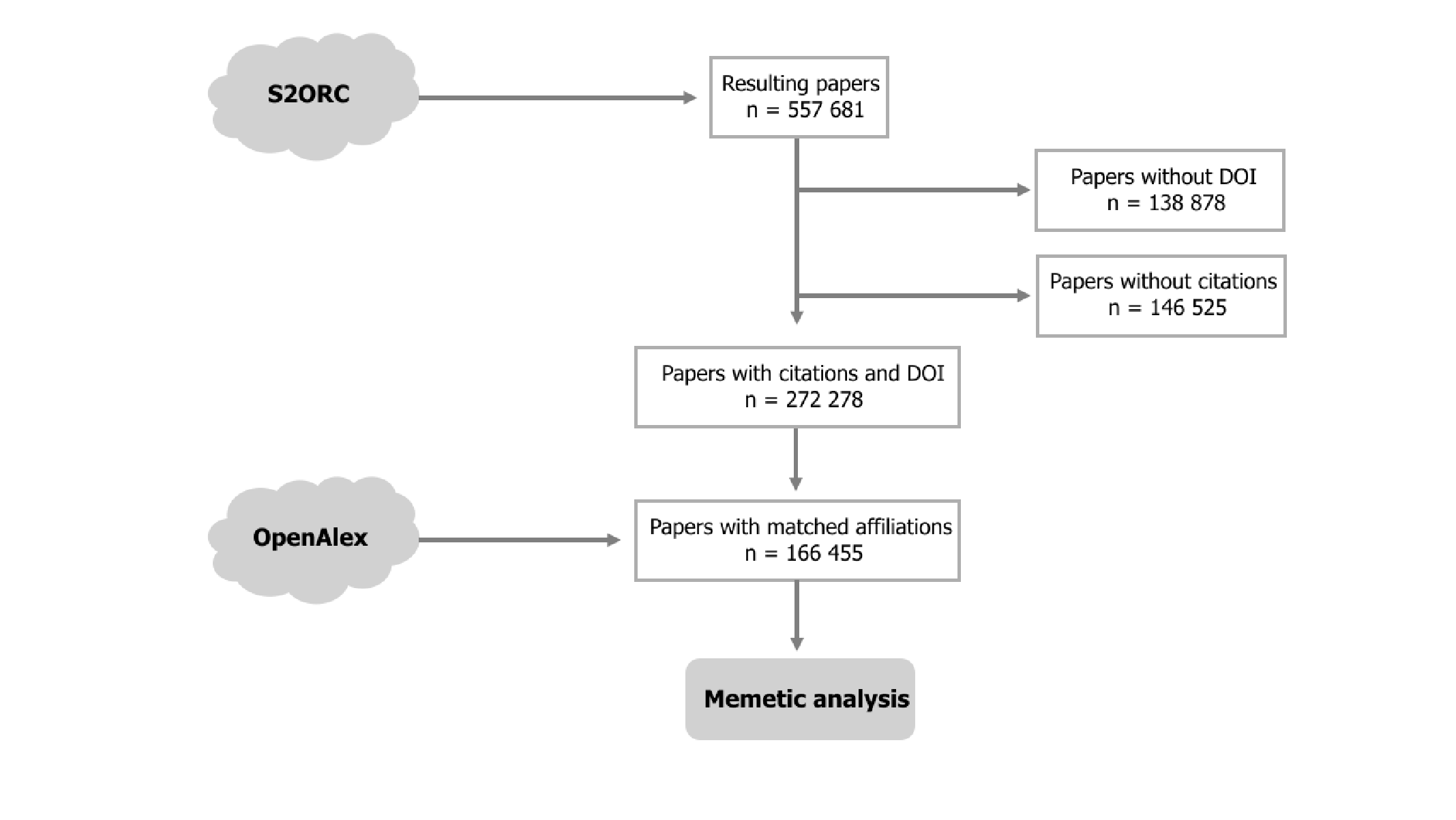}
    \caption{Diagram of pipeline used for gathering and processing papers. }
    \label{fig:data-proc-pipeline}
\end{figure}

The article filtering strategy of our study is based on the work \citet{liu_tracking_2021}. Its authors designed an AI-oriented search query that aims to cover various AI-related topics while maintaining high precision in extracted papers. To achieve these goals, they followed a three-step procedure:
\begin{enumerate}
    \item \textbf{Retrieving AI benchmark records.} The authors downloaded two narrow groups of papers, first from Web of Science (WoS), which include the 'artificial intelligence' term as the topic of the papers, and the second which comes from the 19 top-tier AI-related journals. This way, the relevance of benchmark observations is ensured.
    \item \textbf{Retrieving candidate keywords.} Based on the benchmark records, the authors retrieve both 'Author Keywords' and 'Keywords Plus' to create a wide range of AI-related candidate keywords. Additionally, they remove the general terms (i.e. 'system') and ascertain their meaning by checking online web sources like Wikipedia. This way, the list of 214~candidate keywords is created.
    \item \textbf{Creating 3 keyword groups based on their relevance.} Finally, the authors refine the candidate keywords by using co‑occurrence analysis and Hit Ratio screening.\\
    The first group, called 'core', contains ten terms that yield the highest co-occurrence numbers with the 'artificial intelligence' keyword. The list includes the following terms: \textit{artificial intelligence, neural network, machine learning, expert system, natural language processing, deep learning, reinforcement learning, learning algorithms, supervised learning, intelligent agents}.\\
    In order to validate the remaining 204 terms, the authors calculated the Hit Ratio value for each term, which signifies how many records captured by a candidate keyword are also captured by the 'core' lexical query. If the Hit Ratio was higher than 70\%, then the keyword was added to the 2nd group; if it was between 30\% and 70\%, then it was manually checked if it was used in AI-related paper, and the score below 30\% automatically excluded the candidate keyword.
\end{enumerate}
To ensure the high relevance of AI-related papers in our study, we decided to follow the first filtering strategy, which includes 10 keywords. Additionally, we decided to include a subset of 19 terms that occur in the second and third groups and were deemed significant for the team of three domain (machine learning) experts. This additional list includes the following terms \textit{unsupervised learning, backpropagation learning, backpropagation algorithm, long short term memory, autoencoder, q learning, feedforward net, xgboost, transfer learning, gradient boosting, generative adversarial network, representation learning, random forest, support vector machine, multiclass classification, robot learning, graph learning, naive bayes classification, classification algorithm}. Those filters were used in the first step of the data-gathering pipeline shown in Figure \ref{fig:data-proc-pipeline}.

\section{OpenAlex categories of affiliations treated as Academia}

\setlength{\tabcolsep}{6pt}
\begin{table}[H]
    \centering
    \begin{tabular}{ll}
    \hline
        Category & Most important institutions \\ \hline
         & Tsinghua University  \\ 
        ~ & Shanghai Jiao Tong University  \\ 
        education & Carnegie Mellon University  \\ 
        ~ & Stanford University  \\ 
        ~ & Massachusetts Institute of Technology \\ \hline
        ~ & Chinese Academy of Sciences \\ 
        ~ & French National Centre for Scientific Research \\ 
        government & National Institute of Health \\ 
        ~ & French Institute for Research in Computer Science and Automation \\ 
        ~ & Commonwealth Scientific and Industrial Research Organisation \\ \hline
        ~ & Max Planck Society  \\ 
        ~ & Electric Power Research Institute \\ 
        nonprofit & German Research Centre for Artificial Intelligence \\ 
        ~ & Instituto de Salud Carlos III  \\ 
        ~ & SRI International  \\ \hline
        ~ & Ecole Polytechnique Federale de Lausanne \\ 
        ~ & National Institute of Informatics \\ 
        facility & Electronics and Telecomunications Research Institute \\ 
        ~ & United States Air Force Research Laboratory \\ 
        ~ & Italian Institute of Technology \\ \hline
        ~ & Mayo Clinic \\ 
        ~ & Boston Children’s Hospital \\ 
        healthcare & Brigham and Women’s Hospital \\ 
        ~ & The University of Texas Southwestern Medical Centre \\ 
        ~ & Vanderbilt University Medical Centre \\ \hline
    \end{tabular}
    \caption{Five institutions with the biggest number of publications from each OpenAlex category included by us in the Academia category. Due to overlap in education, government and facility categories we decided to merge them into Academia.}
    \label{tab:annotations}
\end{table}

\section{Annotations}
\label{appendix:Annotations}
%\todo{Tutaj dać link do excela / inaczej pokazanych tych anotacji, dokłądnie opisać proces: memy -> krótkie opisy -> duplikaty -> sposób wyboru klasy. Może wstawić kawałek tabelki?}

The annotation process was conducted for the top memes, defined as the ones that have a meme score value over 0.25, and more than 20 occurrences, by three of domain experts. The pipeline for the creation of the annotations consisted of several steps, described below.

\begin{enumerate}
    \item Extraction of basic parameters: meme score, meme occurrences, and meme phrases.
    \item Providing a short description of more complex, and not obvious memes. The research was focused on providing the descriptions connected to ML, as we know, that used dataset comes from this domain. Providing the titles of ML papers found connected to researched memes.
    \item Providing the flags for duplicate or very similar memes. Most of them were the abbreviations and their elaborations.
    \item Creation of first frequently repeating topics (ML algorithms, Medical terms, Security, Niche ML application, ML concept, and Other), based on the annotation process.
    \item Iterative extraction of another topic from the Other group (Maths, Graphs, Computer Vision, NLP, Data Related).
    \item Revising whole annotations and re-assignment to proper topics.
\end{enumerate}

\newpage
\section{Google, Microsoft, and OpenAI citation count comparison}
\label{appendix:openai}

\begin{figure}[H]
    \centering
    \includegraphics[width=0.9\linewidth]{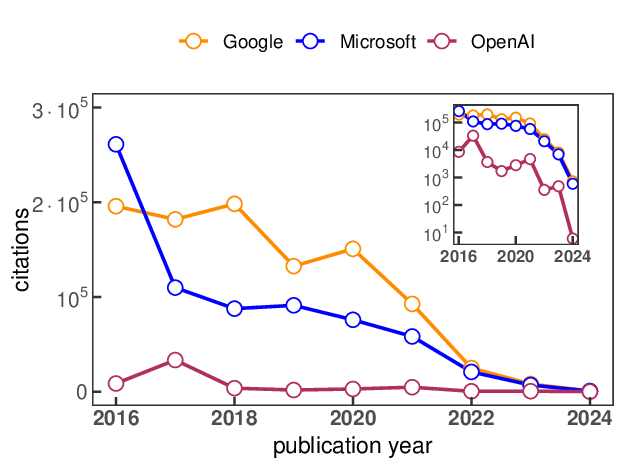}
    \caption{Cumulative number of citations (y-axis) to all papers published in the given year (x-axis) by selected institutions. The inset presents the same data in the log scale. The results show that Microsoft and Google have significantly more total citations than OpenAI. Data: OpenAlex search performed on 23rd Jun 2024.}
    \label{fig:openai-citations}
\end{figure}
\end{document}